\documentclass{article}
\usepackage{geometry}
\usepackage{float}
\usepackage{graphicx}
\usepackage{subfigure}
\usepackage{bbm}
\usepackage{amsmath}
\usepackage{amsfonts}
\usepackage{bm}
\usepackage{graphicx}
\usepackage{enumitem}
\usepackage{float}
\usepackage{authblk}
\usepackage{hyperref}

\title{Some Thoughts on Symbolic Transfer Entropy}
\author[1]{Dian Jin\thanks{Corresponding author} }

\affil[1]{\textit{Research Institute for Intelligent Wearable Systems, The Hong Kong Polytechinic University}}

\begin{document}
\maketitle

    

\begin{abstract}
    Transfer entropy is used to establish a measure of causal relationships between two variables. Symbolic transfer entropy, as an estimation method for transfer entropy, is widely applied due to its robustness against non-stationarity. This paper investigates the embedding dimension parameter in symbolic transfer entropy and proposes optimization methods for high complexity in extreme cases with complex data. Additionally, it offers some perspectives on estimation methods for transfer entropy.
\end{abstract}

\section{Introduction}
Symbolic Transfer Entropy (STE)\cite{staniek2008symbolic}, as a combination of Permutation Entropy (PE) and Transfer Entropy (TE)\cite{schreiber2000measuring}, can quantify the dominant direction of information flow between time series from both identical and non-identical coupled systems. STE measures bivariate information causality, and for multivariate cases, Partial Symbolic Transfer Entropy (PSTE)\cite{papana2013partial} is proposed. MPSTE\cite{9744471} extends PSTE across multiple delay times.

These methods are based on permutation entropy, involving two crucial parameters: delay time $\tau$ and embedding dimension $m$. The initial suggestion for choosing the embedding dimension was $m = 3, \ldots, 7$\cite{bandt2002permutation}. Although smaller $m$ values have been effective in some analyses, such as \cite{li2013parameter}, it was found in \cite{cuesta2018patterns} that using $m$ values from 3 to 9, analyzing EEG records of 4096 samples, temperature records of 480 samples, RR records of 1000 samples, and continuous glucose monitoring records of 280 samples, $m = 9$ provided the highest classification performance even for the shortest signal type. In fact, higher $m$ values often offer better signal classification performance, capturing the underlying signal dynamics more effectively\cite{cuesta2019embedded}. It is recommended to choose the maximum $m$ that satisfies $N > 5m!$\cite{riedl2013practical}.

In certain scenarios, such as in military aircraft, many critical sensors typically have a sampling frequency of $1000\text{ Hz}$. For a 6-hour flight mission, the data volume is $1000 \times 60 \times 60 \times 6 = 2.16 \times 10^7$. \textbf{In the worst-case scenario}, where all sub-patterns of symbolic sequences appear, considering $m^* = \mathop{\arg\max}_{m \in \mathbb{N}^+}\{N > 5m!\}$, we find $m^* = 10$. At this point, $m^*! = 3.6288 \times 10^6$, which imposes a significant computational burden. Additionally, similar patterns may be dispersed, appearing infrequently, contributing minimally to the results. Hence, some optimization can be conducted in this aspect.

\section{Methodology}
STE applies the idea of PE to TE, using a sorting method to ignore the absolute magnitude of elements in the sequence while preserving their relative magnitudes. Since TE calculation is relatively difficult, this method provides an easier estimation of TE while also mitigating the effects of non-stationarity.

Essentially, if ${S}$ is a set consisting of $k$ sequences $\boldsymbol{s}_1,\cdots,\boldsymbol{s}_k$, and ${P}$ is a set of $n$ elements $\boldsymbol{p}_1,\cdots,\boldsymbol{p}_n$, the objective is to find a mapping function $f({S})={P}$ that maps the $k$-dimensional space $\mathcal{S}$ to the $n$-dimensional space $\mathcal{P}$. Generally, $k>>n$.

\begin{figure}[H]
    \centering
    \includegraphics[width=.89\linewidth]{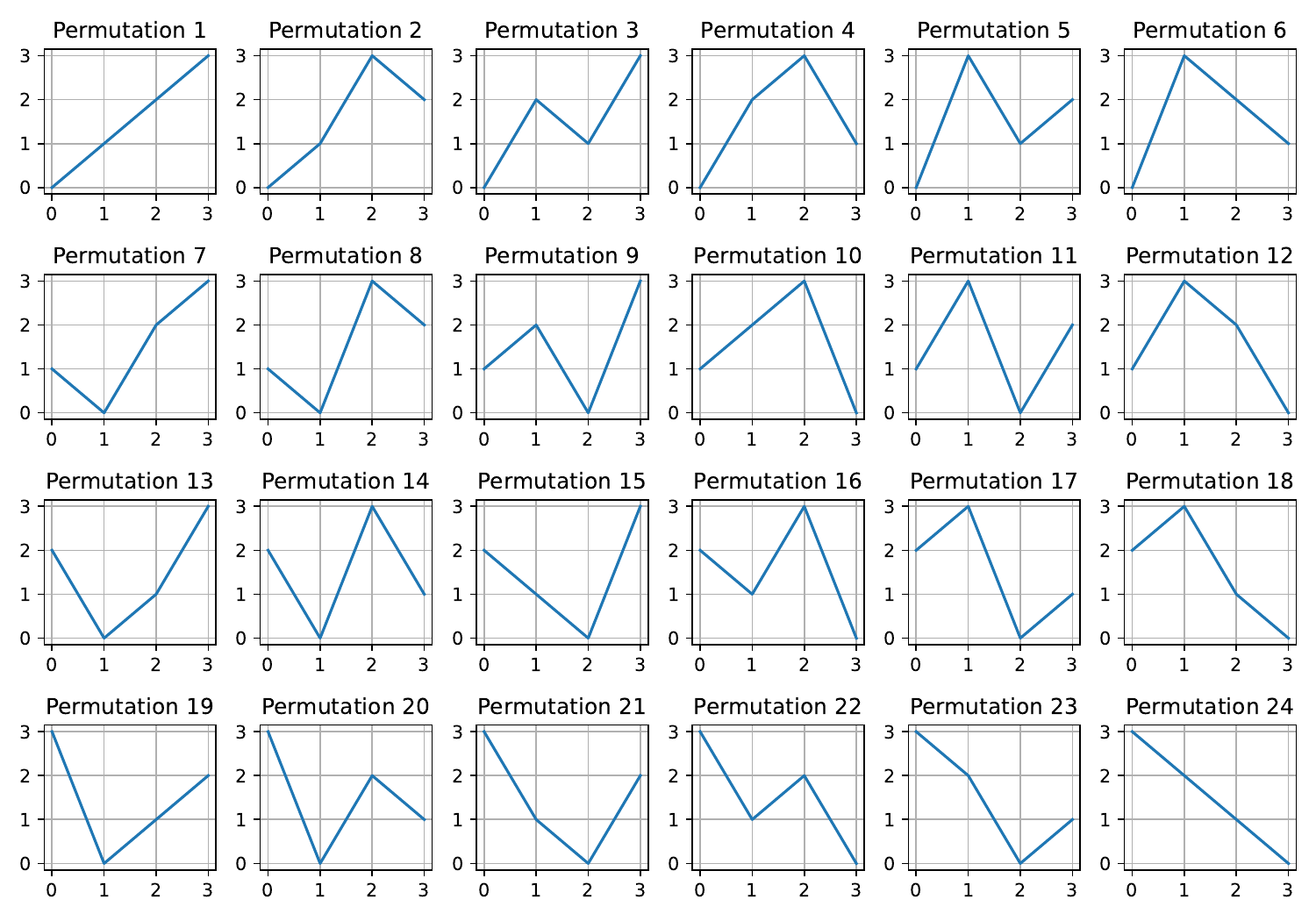}
    \caption{permutations of (0,1,2,3)}
    \label{fig:perm}
\end{figure}

Therefore, STE proposes only one mapping method, and in practice, various mapping methods can be proposed. For different types of data, certain methods may yield better results. STE has a complexity of $O(m!)$. For example, for a sequence with 5 elements, the number of permutations is $m!=4!=24$, as shown in Figure \ref{fig:perm}. This paper considers the optimization of symbolic transfer entropy estimation methods for long, complex time series.

This paper presents several methods to simplify the estimation of transfer entropy\footnote{The code for this paper can be found at \url{https://gitee.com/the-duke/symbolic-transfer-entropy}}.

\subsection{Binning STE}

\begin{figure}[H]
	\centering
	\includegraphics[width=.86\linewidth]{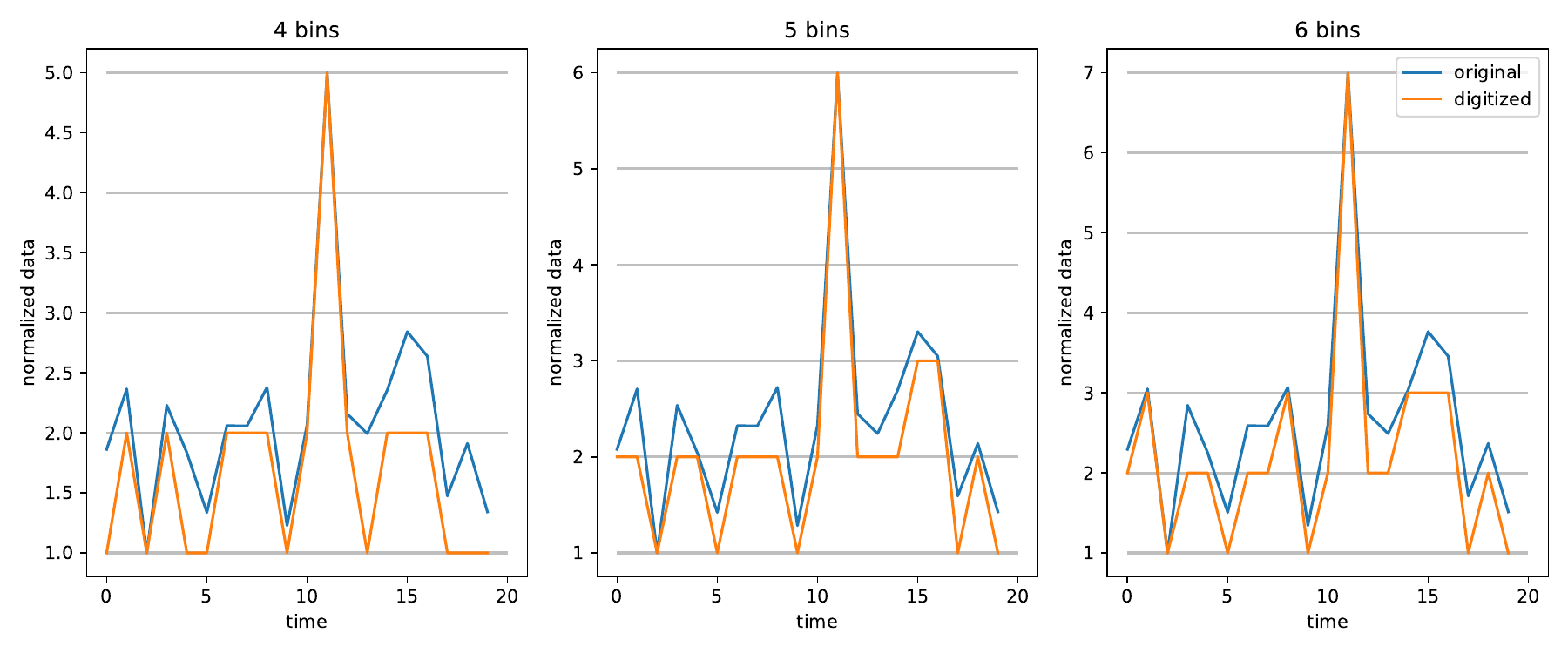}
	\caption{Binning}
	\label{fig:bins}
\end{figure}

A simple idea is to directly bin the subsequences. By distributing the elements of the subsequence into $b$ bins, the number of permutations becomes $b^m$, with a complexity of $O(b^m)$, thus achieving dimensionality reduction, as shown in Figure \ref{fig:bins}. Moreover, this symbolization method can effectively solve the equal value problem mentioned in \cite{vidybida2020calculating}. In fact, when $b=m$, it can evolve into the method proposed in \cite{vidybida2020calculating}.

\subsection{Principal STE}

Another simple idea is to consider only the indices of the largest and smallest groups of values in the sequence elements $\boldsymbol{s}_i$. When considering only one group of extreme values, the number of permutations is $m(m-1)$. When considering $t$ groups of extreme values, the number of permutations becomes $m!/(m-2t)!$. The complexity is $O(m^{2t})$. We consider these groups of extreme values as the principal components of the sequence. Dimensionality reduction is achieved by ignoring secondary information.



\section{Experiment}
\begin{figure}[htbp]
	\centering
	\includegraphics[width=.9\linewidth]{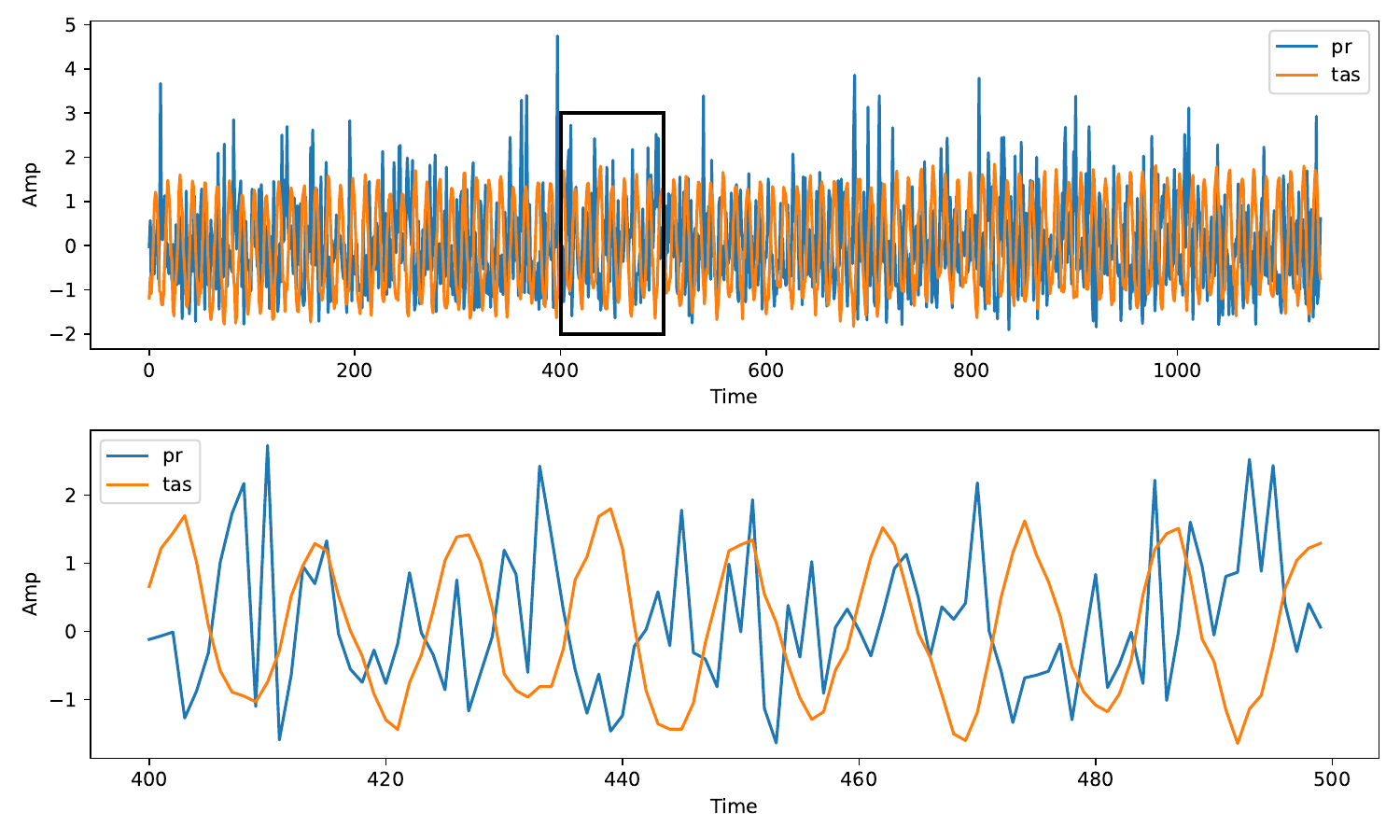}
	\caption{New York precipitation and temperature, with the lower graph showing a magnified view of the upper graph}
	\label{fig:data}
\end{figure}

The dataset chosen for the experiment\footnote{For details on the open-source dataset, see \url{https://data.cig.uw.edu/rocinante/CMIP5/rcp60/GFDL-ESM2M/}} consists of precipitation and temperature data for major global cities. This paper analyzes the data for New York, as shown in Figure \ref{fig:data}.

\begin{figure}[htbp]
	\centering
	\begin{minipage}{0.49\linewidth}
		\includegraphics[width=1\linewidth]{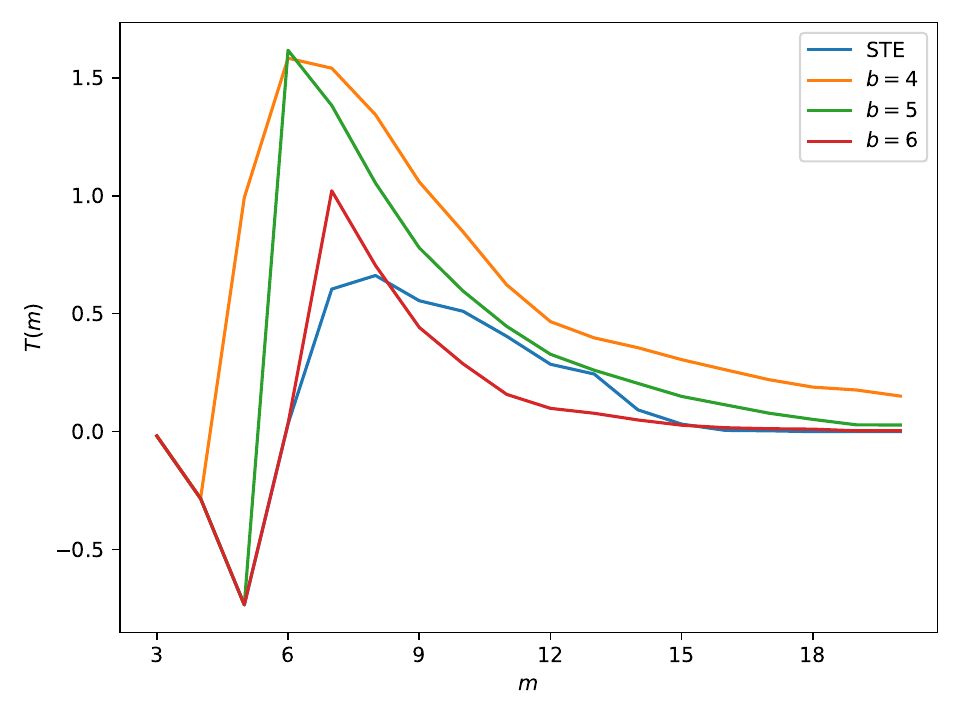}
		\caption{Comparison of Binning STE}
		\label{fig:bTE}
	\end{minipage}
	\begin{minipage}{0.49\linewidth}
		\includegraphics[width=1\linewidth]{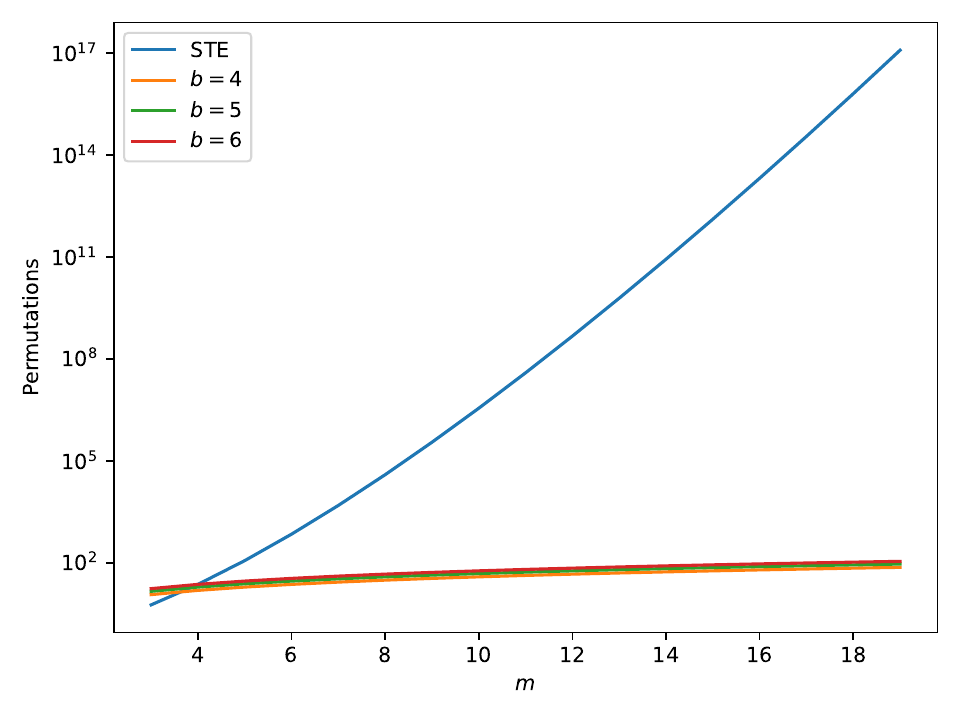}
		\caption{Comparison of permutation numbers for extreme cases in Binning STE}
		\label{fig:bfunc}
	\end{minipage}
\end{figure}

The Transfer Entropy (TE) calculated using the Binning STE method is shown in Figure \ref{fig:bTE}. The complexity is illustrated in Figure \ref{fig:bfunc}. For $b=5,6$, while significantly reducing complexity, it also manages to preserve the original information well (using STE as a reference).

MSE($b$=4): 0.4186908

MSE($b$=5): 0.1874564

MSE($b$=6): 0.0201676

\begin{figure}[htbp]
	\centering
	\begin{minipage}{0.49\linewidth}
		\includegraphics[width=1\linewidth]{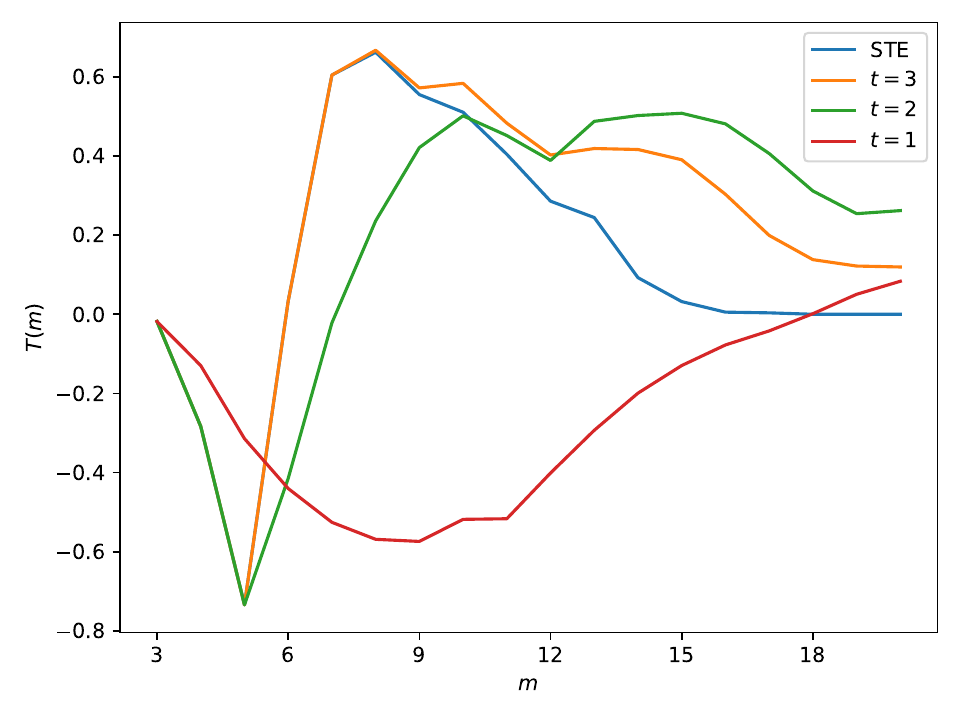}
		\caption{Comparison of Principal STE}
		\label{fig:TE}
	\end{minipage}
	\begin{minipage}{0.49\linewidth}
		\includegraphics[width=1\linewidth]{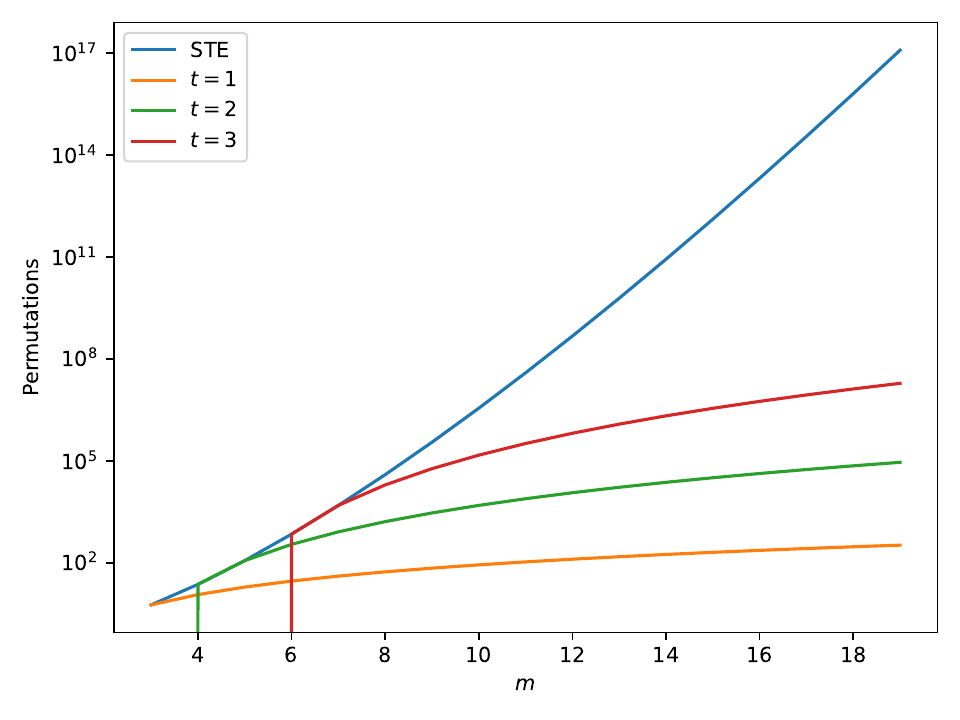}
		\caption{Comparison of permutation numbers for extreme cases in Principal STE}
		\label{fig:functions}
	\end{minipage}
\end{figure}

The Transfer Entropy (TE) calculated using the Principal STE method is shown in Figure \ref{fig:TE}. The complexity is illustrated in Figure \ref{fig:functions}. For $t=2,3$, it manages to preserve the original information well (using STE as a reference). The accuracy is similar.

MSE($t$=1): 0.4045390

MSE($t$=2): 0.1042124

MSE($t$=3): 0.0257761

\section{Conclusion and Future Prospects}
Symbolic Transfer Entropy (STE) symbolizes sequences to estimate transfer entropy. However, when selecting a large $m$ and in extreme cases with numerous permutations, the complexity is high. This paper proposes several methods to simplify the estimation of transfer entropy: the first is Binning STE, which reduces the number of permutations by binning subsequence data; the second is Principal STE, which estimates STE by taking several groups of extreme values from the subsequence, preserving the main information.

The limitation of this study lies in the insufficient size and complexity of the dataset, which fails to demonstrate the optimization effect and may even result in negative optimization. Moreover, the application scenarios are limited to long, complex time series, with optimization being meaningful only when the sequence complexity is relatively high.

\begin{figure}[htbp]
	\centering
	\includegraphics[width=.6\linewidth]{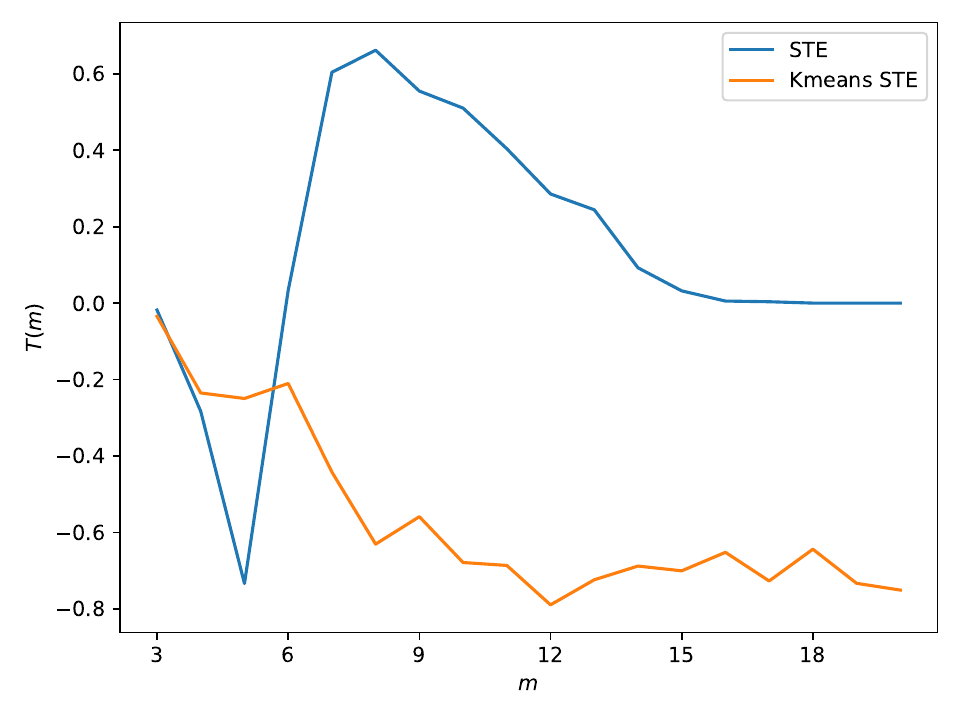}
	\caption{Comparison of K-means STE}
	\label{fig:kmeansTE}
\end{figure}

Additionally, other possible methods include: 1. Discrete wavelet transform, although it generally cannot achieve dimensionality reduction for sequences composed of few sampling points; 2. Traditional dimensionality reduction methods, such as clustering. However, for uniformly distributed samples, the effect may be mediocre: this paper attempted to use the K-means clustering algorithm for dimensionality reduction, with results shown in Figure \ref{fig:kmeansTE}, demonstrating poor performance; 3. The essence of transfer entropy estimation methods is to find similar patterns in the data (differences between $(x_{i+1}|x_i,y_i)$ and $(x_{i+1}|x_i)$ patterns). Perhaps we could start from transfer entropy itself to explore pattern definition methods, both improving estimation efficiency and increasing estimation accuracy.

Regarding trend information in time series data, oscillations or some discretization steps may exist in the system, causing input and output time intervals to not strictly correspond. Perhaps the Dynamic Time Warping (DTW) method \cite{muller2007dynamic} could be used to reduce the impact of this factor.


\bibliographystyle{IEEEtran}
\bibliography{lib}

\end{document}